\documentclass[submission,copyright,creativecommons]{eptcs}
\usepackage[utf8]{inputenc}
\usepackage{amsmath}
\usepackage{amsfonts}
\usepackage{amssymb}
\usepackage{makeidx}
\usepackage{graphicx}
\usepackage{lmodern}
\providecommand{\event}{REF 2015} 
\title{Unifying Theories of Mobile Channels}
\author{Gerard Ekembe Ngondi
\institute{University of York \\ York, UK}
\institute{Department of Computer Science}
\email{gen501@york.ac.uk}
}
\def\titlerunning{Channel Mobility in UTP}
\def\authorrunning{Gerard Ekembe Ngondi}
\usepackage{circus}
\usepackage{csp}
\newtheorem{definition}{Definition}
\newtheorem{theorm}{Theorem}
\newcommand{\healthy}[1]{\hbox{\bfseries\slshape#1}}%
\newcommand{\negbigskip}{\vspace{-1ex}}%
\DeclareMathAlphabet{\mathcal}{OMS}{cmsy}{m}{n}
\providecommand{\urlalt}[2]{\href{#1}{#2}} \providecommand{\doi}[1]{doi:\urlalt{http://dx.doi.org/#1}{#1}}
\begin{document}
\maketitle
\section*{Abstract}\negbigskip
In this paper we present the denotational semantics for channel mobility in the Unifying Theories of Programming (UTP) semantics framework. The basis for the model is the UTP theory of reactive processes (precisely, the UTP semantics for Communicating Sequential Processes (CSP)), which is slightly extended to allow the mobility of channels: the set of actions in which a process is authorised to participate, originally static or constant (set during the process's definition), is now made dynamic or variable: it can change during the process's execution. A channel is thus moved around by communicating it via other channels and then allowing the receiving process to extend its alphabet with the received channel. New healthiness conditions are stated to ensure an appropriate use of mobile channels.

\negbigskip\negbigskip\negbigskip%
\section{Introduction}\label{mobC:intro}\negbigskip
Channel mobility refers to any model or theory for describing systems with a dynamic network configuration also called dynamic (network) or mobile systems. Such systems may be considered as static (network) systems with an additional feature: changing their topology (i.e. the links between their components/processes). In short we may write: \emph{mobile system = static system + function that moves channels}. \\%
The pi-calculus \cite{Mil99} is the most well-known theory concerned with the movement of channels or links: ``links move in a virtual space of linked processes". The semantics for channel mobility are operational and channel mobility is characterised by the communication of names representing the channels. Some denotational models for the pi-calculus \cite{Sta96, Fio96, Hen02} have been given, but the result is not adequate for reasoning about channel mobility directly.  \\%
Other denotational semantics have been attempted: \cite{Broy95, GroSto99} use formalisms based on I/O relations as their basis (c.f. \cite{Broy93}), and \cite{WelBa08, HoaOh08, BialPes09a, BialPes09b, Ros10} are more or less based on CSP \cite{Hoa85, Ros98}. As a general remark, they are all less expressive than the model we present here: either they model only traces, or their modelling technique limits the range of mobile systems which they may serve to model. We discuss them in more detail in section \ref{mobC:relwork}.
\\%
In this paper we present a denotational semantics for channel mobility in UTP \cite{HoaHe98}. The construction of the theory will be based on the UTP theory of CSP (or UTP-CSP -- \cite{HoaHe98, CavWoo06}), which formalises static systems and to which number of modifications will be brought. Following the pi-calculus, channels are communicated as messages through other channels. An immediate consequence is to make the interface of processes (viz. their set of channels) increase or shrink according to the direction (sending or receiving) of the movement. The concept of \emph{dynamic alphabetised trace} (or DAT) will permit us to model dynamic interfaces, and new healthiness conditions will be introduced. \\%
Section \ref{utp:reac} contains a presentation of static networks and their semantics in UTP as reactive processes. Section \ref{mobC:concepts} contains concepts of dynamic networks and their formalisation. The semantics of operators including the operations for moving channels are presented in section \ref{mobC:semX}. We discuss related work in section \ref{mobC:relwork}. Finally, the Conclusion summarises the paper and gives directions for future work.
\negbigskip\negbigskip\negbigskip%
\section{Static Networks - UTP theory of reactive processes}\label{utp:reac}\negbigskip
UTP \cite{HoaHe98} is a formal semantics framework for reasoning about programs, programming theories and the links between theories. The semantics of a program are given by a relation between the initial (undecorated) and final (decorated) observations that can be made of the variables that characterise the program behaviour. Relations are themselves represented as \emph{alphabetised predicates}, i.e.\ predicates of the form $(\alpha P, P)$. $\alpha P$ is called the \emph{alphabet} of the predicate $P$, and determines what variables $P$ may mention. $\alpha P$ may be partitioned into two subsets: $in \alpha P$, which represents the initial observations, and $out \alpha P$, which represents the final observations. \\%
Programming languages and paradigms are formalised as UTP theories. A UTP theory is just a collection of predicates, and consists of three elements: an alphabet, containing only those variables that the predicates of the theory may mention; a \emph{signature}, which contains the operators of the theory, and \emph{healthiness conditions} which are laws constricting the set of legal predicates to those that obey the properties expressed by the conditions. \\%
Healthiness conditions generally have the form: %
$\healthy{NAME} \quad P = f(P)$, %
for some idempotent function $f$ (i.e.\ $f \circ f(x) = f(x)$). $\healthy{NAME}$ stands for the name of the healthiness condition and is also used as an alias for $f$ i.e.\ we write $P = \healthy{NAME}(P)$ and we say that $P$ is $\healthy{NAME}$-healthy. \\%
Specifications are also expressible in UTP, and a theory of refinement permits us to ensure the correctness of a program with regard to a given specification. \\%
A number of programming constructs and paradigms have been formalised using UTP. For example, the theory of \emph{Relations} allows us to specify most of the constructs of sequential programming, including: %
\begin{itemize}
      \item variable declaration: $\mathbf{var}~ x : T$ ; undeclaration: $\mathbf{end}~ x$ ; assignment: $x := e$
      \item termination: $SKIP$ ; conditional choice: $P \lhd b \rhd Q$ ; iteration: $b*P$  %
      \item sequential composition: $P \comp Q$ ; internal choice: $P \intchoice Q$ ; recursion: $\mu X @ F(X)$ %
\end{itemize}  
In what follows we present in slightly greater detail the theory of \emph{Reactive Processes}. %
\negbigskip\negbigskip\negbigskip%
\subsection{Reactive Processes}
The UTP theory of Reactive Processes \cite{HoaHe98, CavWoo06} permits modelling of programs that may interact with their environment. Reactive programs are expressed as processes i.e.\ predicates that allow us to characterise the intermediate states of a program, between initialisation and termination. \\
The interactions of a process are modelled as atomic \emph{events} i.e.\ actions without duration. A process may only engage in given events that are thus said to be \emph{authorised}. Authorised events form a set called the \emph{actions set} or \emph{events alphabet} of the process, and denoted by $\mathcal{A}$ ($\mathcal{A} P$ for a process $P$). \\%
Each occurrence of an event is recorded in order. The resulting sequence is called the \emph{trace} of the process, denoted by the variable $tr: \mathcal{A}\star$. $tr$ gives the trace at the beginning of (i.e.\ prior to) the current observation, $tr'$ gives the trace at the end of the observation. So, $tr' - tr$ gives the trace of the current observation from its start to its end. %
A process or its environment may refuse to engage in an event, for example, when it is engaged in another event. All the events that may be refused constitute the refusal set of the process, denoted by the variable $ref: \power \mathcal{A}$. $ref$ gives events that may be refused during the current observation and $ref'$ give those that may be refused next. \\%
The boolean variable $wait: \boolean$ is used to indicate waiting states. $wait = true$ means that the predecessor is in a waiting state, i.e.\ the process is waiting for its predecessor to terminate, meanwhile it does nothing. When used in conjunction with the boolean variable $ok: \boolean$ it also permits us to indicate termination. The variable $ok$ determines if the process is in a stable state (i.e.\ not making any progress). $ok = false$ means that the current process has not yet started . If $ok = true$ then the process has started and its predecessor is stable. If $ok' = true$ and $wait' = true$ then the process is stable but in an intermediate state. If $ok' = true$ and $wait' = false$ then the process has terminated. If $ok' = false$ then the process is in a non-stable state and the values of other variables  are meaningless: we say that the process \emph{diverges}. \\ %
The variable name $\mathbf{o}$ is often used to stand for all the variables in the set $\{ok, wait, tr, \linebreak ref\}$, also called the \emph{observational variables} of a process. \\%
In summary the alphabet of a reactive process consists of the following: %
\begin{itemize}
    \negbigskip%
    \item $\mathcal{A}$, the set of authorised events ; $tr, tr': \mathcal{A}\star$, the trace ;  $ref, ref': \power \mathcal{A}$, the refusal set %
    \negbigskip%
    \item $ok, ok': \boolean$, stability and termination ; $wait, wait': \boolean$, waiting states %
    \negbigskip%
    \item $v, v'$, other variables %
\end{itemize} %
The above alphabet alone is not enough to characterise reactive processes. Predicates with such an alphabet must also satisfy the following healthiness conditions.  \\[+0.5ex]%
$\healthy{R1} \quad P = P \land tr \leq tr'$ \\
$\healthy{R2} \quad P ~~=~~  \Intchoice_{s} \{P[s,s \cat (tr'- tr) / tr, tr'] | s \in \mathcal{A}\star\}$ \\
$\healthy{R3} \quad P ~~=~~ (II_{CSP} \lhd wait \rhd P)$ \qquad \textbf{where} \\%
$II_{CSP} \circdef (ok' = ok \land wait' = wait 
  \land tr' = tr \land ref' = ref \land v'= v) 
  \lhd ok \rhd (tr \leq tr')$ \\%
$\healthy{R1}$ states that the occurrence of an event cannot be undone viz.\ the trace can only get longer. %
$\healthy{R2}$ states that the initial value of $tr$ may not affect the current observation. %
$\healthy{R3}$ states that a process does nothing when its predecessor has not yet terminated. $II_{CSP}$ is the process that changes nothing. If not started, $ok = false$, then only trace expansion may be observed; otherwise the values of the variables remain unchanged. \\%
A reactive process is one that satisfies all three healthiness conditions above. We also say that it satisfies the healthiness condition $\healthy{R} = \healthy{R1} \circ \healthy{R2} \circ \healthy{R3}$. Note that the order of the composition is irrelevant (c.f. \cite{CavWoo06}). %
A particular model of reactive processes is provided by the CSP process algebra \cite{Hoa85, Ros98} and presented subsequently.
\negbigskip\negbigskip\negbigskip%
\subsection{UTP-CSP processes: syntax and semantics}
UTP-CSP processes (or simply CSP processes below) are reactive processes that obey the following additional healthiness conditions: \\[+0.5ex]%
$\healthy{CSP1} \quad P = P \lhd ok \rhd tr \leq tr'$  \\%
$\healthy{CSP2} \quad P = P \comp J$ \quad \textbf{where} \\%
$J \circdef (ok \implies ok' \land wait' = wait
  \land tr'= tr \land ref' = ref \land v'=v)$ \\%
$\healthy{CSP1}$ states that if a process has not started ($ok = false$) then nothing except for trace expansion can be said about its behaviour. Otherwise the behaviour of the process is determined by its definition. %
$\healthy{CSP2}$ states that a process may always terminate. It characterises the fact that divergence may never be required.  \\%
A CSP process is one that satisfies the healthiness conditions $\healthy{R}$, $\healthy{CSP1}$ and $\healthy{CSP2}$. We also say that it satisfies the healthiness condition $\healthy{CSP} = \healthy{R} \circ \healthy{CSP1} \circ \healthy{CSP2}$. Again, the order of the composition is irrelevant. %
Below we give the semantics of some CSP processes. \\[+1ex]%
\textbf{Assignment.} Denoted by $x := e$, is the process that sets the value of the variable $x$ to $e$ on termination, but does not modify the other variables. It does not interact with the environment, always terminates and never diverges. \\
\indent%
$(x := e) \circdef \healthy{R3} \circ \healthy{CSP1} (ok' \land \lnot wait' \land tr' = tr \land x' = e \land v' = v)  $ \\%
A particular kind of assignment is one that leaves everything unchanged. \\%
\indent%
$II_{CSP} \circdef (ok' = ok \land wait' = wait 
    \land tr' = tr \land ref' = ref \land v'=v) 
    \lhd ok \rhd tr \leq tr' $ \\[+1ex]%
\textbf{Skip.} Denoted by $SKIP$, is the process that refuses to engage in any event, terminates immediately and does not diverge. It is special kind of $II_{CSP}$. \\%
\indent%
$SKIP \circdef \exists ref \circspot II_{CSP}$ \\[+1ex]
\textbf{Stop.} Denoted by $STOP$, is the process that is unable to interact with its environment. It is always in a waiting state. It represents deadlock. \\%
\indent%
$STOP \circdef \healthy{R} (wait := true)$ \\[+1ex]
\textbf{Chaos.} Denoted by $CHAOS$, is the process with the most non-deterministic behaviour viz.\ the worst possible reactive process. It represents divergence.  \\%
\indent%
$CHAOS \circdef \healthy{R} (true)$ \\[+1ex]
\textbf{Sequential composition.} Denoted by $P \comp Q$, is the process that first behaves like $P$, and if $P$ terminates, then behaves like $Q$.  %
\negbigskip%
\begin{align*}
& P(\mathbf{o'}, v') \comp Q(\mathbf{o}, v) \circdef \exists \mathbf{o_{0}}, v_{0} \circspot P(\mathbf{o_{0}}, v_{0}) \comp Q(\mathbf{o_{0}}, v_{0})  
& \text{, provided}~~ out \alpha P = in \alpha ' Q = \{\mathbf{o'}, v'\}
\end{align*}
%
%
%
\textbf{Internal choice.} Denoted by $P \intchoice Q$, is the process that behaves either like $P$ or like $Q$, where the choice cannot be controlled by the environment. \\%
\indent%
$P \intchoice Q \circdef P \lor Q$ \\[+1ex]
\textbf{External choice.} Denoted by $P \extchoice Q$, is the process that behaves like $P$ or $Q$, where the choice is controlled by the environment. \\%
\indent%
$P \extchoice Q \circdef \healthy{CSP2} \big((P \land Q) \lhd STOP \rhd (P \lor Q) \big)$ \\[+1ex]
The definition states that if no interaction is performed and termination has not occurred i.e.\ $STOP = true$ then the observation must be agreed by both $P$ and $Q$. Otherwise the behaviour will be either that of $P$ or $Q$ depending on which one the environment chose to interact with. Note that $STOP$ may appear here in a conditional statement because in the end, $STOP$ is but a predicate, that is, its presence occurs as the result of trying to reduce the expression defining $P \extchoice Q$. \\[+1ex]%
\textbf{Prefix.} Denoted by $a \then P$, is the process that engages in action $a$ and then behaves like process $P$. First consider the following function.%
\begin{definition}[$\Phi$]\label{df:phi-do}
$\Phi \circdef \healthy{R} \circ and_{B} = and_{B} \circ \healthy{R}$ \\%
\textbf{where} $B \circdef (tr' = tr \land wait') \lor tr < tr'$ \\%
\textbf{and} $and_{B} \circdef \lambda X \circspot B \land X$,\footnote{We use a lambda notation in place of $and_{B}(X) \circdef B \land X$} \emph{where} $X$ denotes any predicate of a given UTP theory.
\end{definition}
$and_{B}$ imposes the condition $B$ on any predicate $X$: if $X$ is waiting for an event to occur then it may not modify the trace, otherwise the only possible observation is that the trace has changed. $\Phi$ makes the predicate $X$ into a reactive process and permits us to determine the values of the other variables when $wait' = true$. \\%
The occurrence of an event $a$ is defined as the process denoted by $do_{\mathcal{A}}(a)$. It never refuses to engage in $a$ whilst it is waiting for $a$ to occur. Following its occurrence, $a$ is recorded in the trace. \\%
\indent%
$do_{\mathcal{A}}(a) \circdef \Phi (a \notin ref' \lhd wait' \rhd tr' = tr \cat \langle a \rangle)$ \\[+1ex]%
As an abbreviation, we may just write $a$ instead of $do_{\mathcal{A}}(a)$. The process denoted by $a \then SKIP$ first engages in $a$ and then terminates successfully. \\%
\indent%
$a \then SKIP \circdef \healthy{CSP1}(ok' \land do_{\mathcal{A}}(a))$ \\[+1ex]%
The process denoted by $a \then P$ first engages in event $a$ and then behaves like some process $P$.\\%
\indent%
$a \then P \circdef a \then SKIP \comp P$ \\[+1ex]%
%
%
\textbf{Communication.} A particular type of interaction between processes is the passing or communication of messages. A \emph{communication} is the sending or the reception of a message via a channel (the communication medium). A communication event is represented by a pair $ch.e$ where $ch$ denotes the (name of the) channel used for the communication, and $e$ any message that may be sent through $ch$. The set of all the channels that a process may use is called its \emph{interface}; it will be denoted by $\mathcal{I} \circdef \{ch | \exists m \circspot ch.m \in \mathcal{A}\}$. \\ %
The occurrence of a communication event $ch.e$ is just the process $do_{\mathcal{A}}(ch.e)$. As earlier we may just write $ch.e$. The definitions below correspond to a synchronous model of communication. \\%
\indent%
$ch.e \circdef do_{\mathcal{A}}(ch.e)$ \\[+1ex]
The input prefix $ch?x \then P$ receives a message from $ch$, assigns it to the variable $x$ and then behaves like process $P$. \\%
\indent%
$ch?x \then P \circdef {\Extchoice}_{e \in Msg} ch.e \comp x := e \comp P$ \\[+1ex]
The output prefix $ch!y \then P$ is the process that is willing to output the value of the variable $y$ through channel $ch$ and then behaves like $P$. \\%
\indent%
$ch!y \then P \circdef ch.y \comp P$ \\[+1ex]
\textbf{Parallel composition.} Denoted by $P \parallel Q$, is the process that behaves like both $P$ and $Q$ and terminates when both have terminated. $P$ and $Q$ may not share any variable other than the observational variables ($ok, wait, ...$). $P$ and $Q$ modify separate copies of the shared observational variables that are then merged at the end using the merge predicate $M$, as defined below. 
\negbigskip%
\begin{align*}
&\mathcal{A}(P \parallel Q) \circdef \mathcal{A} P \cup \mathcal{A} Q \\[-1ex]%
&P \parallel Q \circdef P(\mathbf{o}, 1.\mathbf{o'}) \land Q(\mathbf{o}, 2.\mathbf{o'}) \comp M(1.\mathbf{o}, 2.\mathbf{o}, \mathbf{o'}) \\[-1ex]%
&M \circdef \left( 
  \begin{aligned}
    &ok' = (1.ok \land 2.ok) \land  \\[-1ex]%
    &wait' = (1.wait \lor 2.wait) \land  \\[-1ex]%
    &ref' = (1.ref \cup 2.ref) \land  \\[-1ex]%
    &(tr' - tr) = \big((1.tr - tr) \parallel (2.tr - tr) \big)
  \end{aligned} \right) \comp SKIP %
\end{align*} %
Upon termination the final trace is given by the trace merge function defined subsequently.
Let $s$ and $t$ be two traces. Let $E(s)$ denote the set of events in $s$. Let $a, b, c, d$ be (pairwise distinct) events such that: $\{a, b\} \notin E(s) \cap E(t)$, $\{c, d\} \in E(s) \cap E(t)$. Then we may define the trace merge for parallel composition by case as follows: 
\negbigskip%
\begin{align*}
& s \parallel t \circdef t \parallel s 
\qquad \langle \rangle \parallel \langle \rangle \circdef \{\langle \rangle\} 
\qquad
\langle \rangle \parallel \langle c \rangle \circdef \{ \} 
\qquad
\langle \rangle \parallel \langle a \rangle \circdef \{\langle a \rangle\} 
\\[-1ex]%
& \langle c \rangle \cat x \parallel \langle c \rangle \cat y \circdef \{\langle c \rangle \cat u | u \in x \parallel y\}  \\[-1ex]%
& \langle a \rangle \cat x \parallel \langle c \rangle \cat y \circdef \{\langle a \rangle \cat u | u \in x \parallel \langle c \rangle \cat y\} \\[-1ex]%
& \langle c \rangle \cat x \parallel \langle d \rangle \cat y \circdef \{\}  \\[-1ex]%
& \langle a \rangle \cat x \parallel \langle b \rangle \cat y \circdef \{\langle a \rangle \cat u | u \in x \parallel \langle b \rangle \cat y\} \cup \{\langle b \rangle \cat u | u \in x \parallel \langle a \rangle \cat y\}  %
\end{align*}
CSP processes only permit the representation of static systems i.e.\ systems whose network configuration does not change during their activation. For that reason the model for CSP processes presented so far will be referred to as the \emph{static model}. In order to represent mobile systems, number of changes need to be provided to that static model; they are presented subsequently.
\negbigskip\negbigskip\negbigskip%
\section{Dynamic (Network) Systems - Concepts and their Formalisation} \label{mobC:concepts}\negbigskip
Process networks whose configuration/topology may change throughout their activation are called dynamic networks or mobile systems. \emph{Channel mobility} is the name of the corresponding paradigm. The observation of a dynamic system may be divided according to its different topologies.
\negbigskip%
\begin{definition}[Snapshot]\label{df:snapshot}
A \emph{snapshot} is a period of fixed network topology for a mobile process. It determines the behaviour of the mobile process for the duration of that period. Hence, for a mobile process, no two consecutive snapshots may be identical (viz.\ describe the same topology). The overall behaviour of the process may be obtained by concatenating all the snapshots of the process in their order. For comparison, there is only one such snapshot for any non-mobile/static process.
\end{definition}
Processes are connected via links or channels through which they may communicate. Consider a network of three processes $P$, $Q$ and $R$ connected as shown by \emph{fig.1(left)}. Another possible topology for such a network may be obtained by removing the link $ch1$ between $P$ and $Q$ and using it to connect $P$ and $R$ instead, as shown by \emph{fig.1(right)}. Note that it is not properly the channel that moves, but its ends. Hence it would be more appropriate to talk of the mobility of channel ends, and this is what should be understood in the subsequent paragraphs. %
\begin{figure} 
\centering
\includegraphics[scale=0.3825]{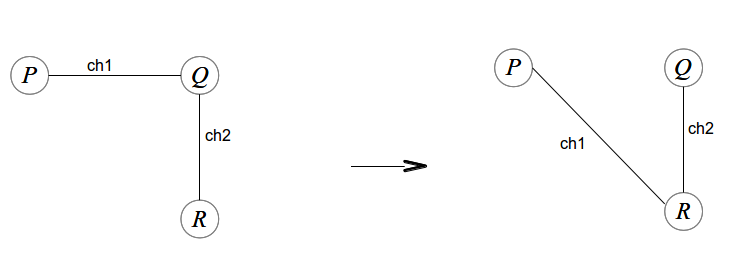}
\negbigskip%
\caption{{Channel Mobility with 3 processes.}\label{fig:mobC3proc} 
{(left)Before the migration of $ch1$.}\label{fig:mobC3proc:bef}
{(right)After the migration of $ch1$}\label{fig:mobC3proc:aft}
}
\end{figure}
In this section we present the necessary changes to the static model (section \ref{utp:reac}) that enable us to give semantics to channel mobility. The mobility model has three main characteristics:
\negbigskip%
\begin{itemize}
\item channels are `localised' in alphabets or more precisely, in interfaces. Hence the mobility of a channel is from one interface to another.
\negbigskip%
\item channels may be communicated as messages amongst processes (hence we need a new representation for channels);
\negbigskip%
\item the interface of processes may change as a consequence of channel mobility.
\end{itemize}%
\textbf{Channel names.} In the static model of CSP processes, channel names are just \emph{logical identifiers}.\footnote{`logical' in the sense that a channel name $ch \in \mathcal{I}$ represents/models a logical concept viz.\ `the occurrence of a communication on the channel named $ch$', and not the channel itself.} For channel mobility channels must rather be modelled explicitly, as data elements: they will also be represented by channel names. How this new set of names relates to the one from the static model, i.e.\ the interface $\mathcal{I} = \{ch | \exists e \circspot ch.e \in \mathcal{A}\}$, is shown below. \\%
Let $Chans$ denote the set of channels that a process may use for its communications. The names in the set $Chans$ represent actual objects or entities, similar to natural numbers. %
All such names must also belong to the interface of the process as defined in the static model i.e.\ $Chans = \mathcal{I}$, for every process.  \\%
In order to bring as little change as possible to the static model, and to keep reasoning about static and dynamic aspects of a process's behaviour separate, we will maintain elements from the static model whenever possible, and add new elements specifically for mobility. \\[+1ex]%
\textbf{Mobile channels.} For the purpose of a static-dynamic dichotomy, $Chans$ will contain static channels only. So we define a set of mobile channels only, denoted by $MCh$. The two sets must be disjoint: $Chans \cap MCh = \{\}$. This will notably ensure that channels in $Chans$ may not be moved. \\%
Channel mobility works with the assumption that a process may receive new channels, i.e.\ channels it did not previously own. We will denote by $mChans: \power MCh$ the variable that may contain such channels when they have been acquired.
\begin{definition} [$MCh$, $mChans$]
We assume a set of mobile channels, denoted by $MCh$. Then: \\
$mChans, mChans': \power MCh$, is the variable that contains the set of channels that have been acquired before the current observation, and are hence authorised. $mChans'$ contains the channels that may be authorised next.
\end{definition}
\negbigskip\negbigskip%
\begin{definition} [Ownership]
A process \emph{owns} a mobile channel $mc$ if and only if $mc \in mChans$.
\end{definition}
\textbf{Events of mobile channels.} In the static model the set $Chans$ is not represented. The interface of a process may be obtained only from its actions set $\mathcal{A}$. With channel mobility, on the contrary, we start with the channels since they are the ones that may move, and then we obtain the corresponding set of events. Hence, we define the set $MCev$ that contains events related with mobile channels only, i.e.\ events of the form $c.m$ where $c \in MCh$.
\begin{definition} [$MCev$]
Let $MCev$ denote the set of events obtained from $MCh$. Then: \\
$MCev \circdef \{ch.e | ch \in MCh\} $. 
\end{definition}
\textbf{Dynamic traces.} Let $mtr$ denote the trace associated with `acquired' mobile channels viz.\ those in $mChans$. The value of $mChans$ at a given time defines which events may be recorded at that time; at different times, $mChans$ may have different values: $mtr$ must reflect such changes. \\%
Whilst in the static model the type of $tr$, $\mathcal{A} \star$, guarantees that only actions in $\mathcal{A}$ may be recorded, to provide the same guarantee in the context of channel mobility by adopting the typing approach of the static model would require that the type of $mtr$ changes whenever $mChans$ takes a new value. This is a problem of dynamic typing that may be solved as follows. \\%
First, we recall that the `type' of a variable determines the values that the variable may take. For the type to change over time simply means that the corresponding set of possible values changes over time. Hence, dynamic typing may be modelled by employing a static type defining every possible values (a sort of default set) and then placing restrictions on that default set where necessary within the process's definition. \\%
In our case, we may define a static type for $mtr$, $(MCev)\star$. This would mean that any event in $MCev$ may be recorded, which is too large. We now need to enforce the condition that only events associated with channels already acquired (i.e.\ in $mChans$) may be recorded. For that purpose, we need to keep the history of successive interfaces i.e.\ the history of the value of $mChans$. We could then ensure that at a given time, the event recorded in $mtr$ belongs to the set value of $mChans$ at that time. That is, at a given observation time $k$, we must record both the value of $mChans$, say $mChans_k$, together with the event, say $e$, and ensure that $e \in mChans_k$. We thus introduce the notion of \emph{dynamic alphabetised trace}. \\%
In any snapshot, the value of $mChans$ is fixed, and differs between any two consecutive snapshots. If we associate the value of $mChans$ within a single snapshot with a valid trace for that snapshot, we obtain an \emph{alphabetised trace}. And if we combine the alphabetised traces of every snapshot into a single trace, we obtain a \emph{dynamic alphabetised trace}. For simplicity however, we associate the occurrence of every event with the valid dynamic interface at the time of the observation -- this permits us to disregard snapshots.
\begin{definition}[DAT]
A \emph{dynamic alphabetised trace} or \emph{DAT} is any trace of the form \linebreak$\langle ..., (s, e), ...\rangle$ where $s$ is the valid dynamic interface (viz.\ given by $mChans$) at the time of the observation, and $e$ is the event recorded at that time. 
\end{definition}
The null event will be denoted by $nil$. For DATs it is more convenient than an empty space. In particular, there may be many events of the form $(s, nil)$. Every trace must contain at least one such event. \\%
In our construction so far, $tr$ and $mtr$ have been considered as non-alphabetised. In particular, we have introduced $tr$ only in relation with the static CSP model. In what follows, we should work with the alphabetised versions only. For a mobile process, the overall trace will be denoted $dtr$.\footnote{The variable name $dtr$ is used here mainly for readability, to keep separate static and mobile CSP theories. The name dichotomy is \emph{not} necessary, especially when discussing the links between the two theories. The discussion of such links is, however, not the object of this paper.} It should contain elements from both $str$ and $mtr$, where $str$ will stand for the (alphabetised) trace relating to static channels exclusively. Their respective relation to $dtr$ is obvious: $str = dtr \filter \mathcal{A}$ and $mtr = dtr \filter MCev$, but we shall keep using them informally for the sake of the presentation. 
\begin{definition}[Trace of a mobile process]
Let $\Sigma$ denote the actions set for mobile processes, then $\Sigma \circdef \{nil\} \cup \mathcal{A} \cup MCev$. \\%
$dtr, ~dtr': (\power \Sigma \times \Sigma) \star$, is the dynamic alphabetised trace of mobile processes. 
\end{definition}%
We use the following two projections to select each component of an element in a DAT trace: %
$ \pi_{1} (s, e) = s,~~ \pi_{2} (s, e) = e$. %
We override them to get also the first and second component of all elements in a trace, respectively. Let $k \in \{1, 2\}$, then %
\negbigskip%
\[
\pi_{k} (\langle (s, e) \rangle) = \langle \pi_{k} (s, e) \rangle \\%
\pi_{k} (head~dtr \cat tail~dtr) = \pi_{k} (head~dtr) \cat \pi_{k} (tail~dtr)\\%
\]%
We note here the relation between the dynamic interface and the trace of a mobile process. The initial value of the dynamic interface may be given by: $mChans = last~\pi_{1}(mtr)$. This relationship makes the presence of $mChans$ into the alphabet appear redundant. However, $mChans$ is justified by construction in the following sense: it is not the trace that determines the value of $mChans$, but the opposite. \\%
We now give a more formal characterisation of the notion of a snapshot defined earlier. We may say that a process has a static network (or fixed network topology -- see Definition \ref{df:snapshot}) when its interface is the same whatever the elements of its DAT. Formally: $provided~~ \#mtr' \geq 2,$ \\%
\indent%
$\healthy{SN} \quad P = P \land \forall (s_1, e_1), (s_2, e_2) \in mtr' |(s_1, e_1) \cat (s_2, e_2) \in mtr' \circspot s_1 = s_2$ \\%
A process must have at least two distinct snapshots (viz.\ must be the concatenation of at least two distinct $\healthy{SN}$ processes) to be considered of having a dynamic topology. In other words, at least two consecutive elements of its trace must have separate interfaces. Formally: $provided~~ \#mtr' \geq 2,$ \\%
\indent%
$\healthy{DN} \quad P = P \land \exists (s_1, e_1), (s_2, e_2) \in mtr' |(s_1, e_1) \cat (s_2, e_2) \in mtr' \circspot s_1 - s_2 \neq \{\}$ \\%
The guarantee that \emph{a process may engage only in channels that it already owns} yields the following healthiness condition: \\%
\indent%
$\healthy{MC1} \quad P ~~=~~ P \land (~ \forall s : \power MCh, e: MCev \circspot \langle(s, e)\rangle \in mtr' \implies e \in s ~) $ \\%
Literally, $\healthy{MC1}$ states that every event $e$ that is recorded must belong to the dynamic interface $s$ (the associated events alphabet) valid at the time of the record. \\%
DATs lead us to reconsider the healthiness condition $\healthy{R2}$. In effect, $\healthy{R2}$ is meant to hold for the events history only, not for other types of history. The application of $\healthy{R2}$ to mobile processes is called $\healthy{R2M}$, given below. %
\negbigskip\negbigskip%
\[\begin{aligned}
\healthy{R2M} \quad P ~~=~~  \Intchoice \{P[s,& \big(s \cat (mtr'- mtr)\big) ~/~ mtr, mtr')] | s \in \Sigma\star \land \\%
&\pi_{2}(s) = \pi_{2}(mtr) \land \pi_{2}\big(s \cat (mtr'- mtr))\big = \pi_{2}(mtr') \} 
\end{aligned}\]%
Literally, $\healthy{R2M}$ allows replacing initial events without regard for their related interface. An immediate consequence is the possibility of associating an interface with an invalid event (a communication whose channel is not in the interface), thus violating $\healthy{MC1}$. This is undesirable as it means that we are defining too many unhealthy behaviour, a future problem for verification. A naive solution would be to allow changing the initial interface history also, hence we could have matching substitute pairs. However, whilst it may be possible to change the initial events history for an arbitrary history, such may not be the case for the interface history. \\%
We may better discuss the consequences of such a change by means of an illustration. \\%
Let $\mathcal{I}1$ be the initial interface of a process, and suppose that we know the interface $\mathcal{I}2$ of the next adjacent snapshot. Let $\mathcal{J}$ be an interface and the substitute for $\mathcal{I}1$.  
\begin{itemize}
\item For $\mathcal{I}1 = \{ch1, ch2\}$, $\mathcal{I}2 = \{ch2\}$, and $\mathcal{J} = \mathcal{I}2$, the substitution is clearly undesirable since it denies the movement of $ch1$.  If instead $\mathcal{J} = \{\}$, then the substitution supposes the mobility of $\{ch2\}$ which would contradict with the definition.%
\negbigskip%
\item Now let $\mathcal{I}1 = \{ch1\}$, $\mathcal{I}2 = \{ch1, ch2\}$, and $\mathcal{J} = \mathcal{I}2$. Again, such a substitution cancels the movement of $ch1$. 
\end{itemize}%
In sum, any substitution of the initial interface history for a distinct interface \emph{may have} a mobility effect, which is unhealthy. \\%
By definition, the mobility of a single channel (or of many together) induces a snapshot dichotomy between the topology before the movement and the one after. %
Clearly, the interface at the end of the previous snapshot is the same as the one at the beginning of the following snapshot. %
Thus, any substitution of the initial trace which conserves the value of $mChans = last~\pi_{1}(mtr)$ is valid, otherwise it is unhealthy. \\%
In sum, \emph{the value of the interface at the end of the previous observation must be the initial value at the beginning of the current one}. This yields the following healthiness condition. \\%
\indent%
$\healthy{MC3} \quad P = \Intchoice \{P[s, \big(s \cat (mtr'- mtr)\big) ~/~ mtr, mtr')] | s \in \Sigma\star \land last~\pi_{1}(s) = first~\pi_{1}(mtr')\} $ \\%
$\healthy{MC3}$ translates the idea that the dynamic interface is always fixed (or completely determined) before entering a new snapshot, and that it is the previous snapshot (process) that fixes it. Also notice that $\healthy{MC3}$ implies $\healthy{R2M}$. \\[+1ex]
\textbf{Refusals.} In the basic model, $ref: \power \mathcal{A}$ contains events in which a process may refuse to engage, although they are authorised. The type of $ref$, $\mathcal{A}\star$, guarantees that a process may only refuse authorised events i.e.\ \emph{a process cannot refuse events that it does not own}. We will denote by $dref$ the refusals set for mobile processes, and by $sref$ and $mref$ its static and mobility components. With channel mobility, the type of $mref$ would need to follow the changes of the dynamic interface, so we would once again face a problem of dynamic typing. As earlier, we may solve it by considering a static type for $mref$, and then impose a restriction on the events that may be refused, by means of a healthiness condition. The static type for $mref$ will be $\power MCev$. The healthiness condition expressing that only owned events may be refused is given below. \\%
\indent%
$\healthy{MC2} \quad P ~~=~~ P \land mref' \subseteq last~\pi_{1}(mtr')$  \\%
Literally, $\healthy{MC2}$ states that the refusal must always be a subset of the event alphabet whose value is determined by $mChans = last~\pi_{1}(mtr)$. \\[+1ex]
\textbf{In summary.} We have presented in this section the fundamental concepts of channel mobility and how they may be formalised in UTP. The formalisation has been based on static UTP-CSP and shows quite clearly, if this was not clear enough, that channel mobility is altogether a new paradigm. %
Three new healthiness conditions have been introduced, and a new trace model has been defined which aggregates interface history to the original events history. 
We have chosen to pair together the elements of both histories the consequence of which is the introduction of a null event $nil$. Another approach is possible where the elements from each history are recorded separately. %
The choice of either model is likely a matter of taste. %
%
\negbigskip\negbigskip\negbigskip%
\section{The Semantics}\label{mobC:semX}\negbigskip
In this section we present the denotational semantics of channel mobility. As every UTP theory, it must have three elements: an alphabet, a signature and healthiness conditions. In section \ref{mobC:concepts} we have introduced alphabet elements as well as some healthiness conditions. In this section we put them together to define what a mobile process is. The operators are defined afterwards. The highlight of this section is the semantics of the operation that may change the interface of a process during its activation.
\begin{definition} [Mobile processes]
A mobile process is one that satisfies the healthiness conditions $\healthy{R1}$, $\healthy{R2M}$,\footnote{Although redundant with $\healthy{MC3}$, the choice of leaving $\healthy{R2M}$ is to make obvious that mobile processes are also reactive (CSP) processes.} $\healthy{R3}$, $\healthy{CSP1}$, $\healthy{CSP2}$, $\healthy{MC1}$, $\healthy{MC2}$, and $\healthy{MC3}$, and has an alphabet consisting of the following: 
  \begin{itemize}
  \negbigskip%
  \item $\mathcal{A}$, the set of static events in which it can potentially engage; the events in this set may not move. %
  \negbigskip%
  \item $MCh$, the set of mobile channels that can potentially be moved in (acquired) or moved out (released) during activation. %
  \negbigskip%
  \item $MCev$, the set of events whose channels are in $MCh$. %
  \negbigskip%
  \item $dtr, dtr': (\power \Sigma \times \Sigma) \star$, (where $\Sigma = \{nil\} \cup A \cup MCev$,) the trace. %
  \negbigskip%
  \item $dref, dref': \power \Sigma$, the refusals set. %
  \negbigskip%
  \item $ok, ok': \boolean$. %
  \negbigskip%
  \item $wait, wait': \boolean$. %
  \negbigskip%
  \item $v, v'$. %
  \end{itemize}\negbigskip%
For simplicity, we may introduce the following variables that can be calculated from those above:
 \begin{itemize}
 \negbigskip%
 \item $iface, iface': MCh \star$, the interface history, the trace obtained by selecting only the first element of pairs $(s, e) \in mtr$ (resp. $mtr'$).
 \negbigskip%
 \item $evt, evt': (\mathcal{A} \cup MCev) \star$, the events history, the trace obtained by selecting only the second element of pairs $(s, e) \in mtr$ (resp. $mtr'$).
 \negbigskip%
 \item $mChans, mChans': MCh$,  the last element of the interface history $iface$ (resp. $iface'$). 
 \negbigskip%
 \item $mEv, mEv': \power MCev$, the set that contains the events whose channels are in $mChans$ (resp. $mChans'$). %
 \negbigskip%
 \item $mtr, mtr': (\power MCev \times MCev) \star$, the partition of the trace $dtr$ (resp. $dtr'$) restricted to mobile channels.
 \negbigskip%
 \item $mref, mref': \power MCev$, the subset of the refusals set $dref$ (resp. $dref'$) restricted to mobile channels.
 \end{itemize}
\end{definition}
\negbigskip\negbigskip\negbigskip%
\subsection{Some mobile processes}\negbigskip%
The definitions below apply exclusively to mobile processes. The notation $\healthy{NAME12..n} = \healthy{NAME1} \circ \healthy{NAME2} \circ .. \circ \healthy{NAMEn}$ will be used as short-hand for naming composite healthiness conditions. The order of the composition does not matter. \\[+1ex]
\textbf{Assignment (2).} The definition of assignment follows the earlier one, except that it must be made healthy. \\%
\indent%
$x := e \circdef \healthy{MC123} \circ \healthy{R3} \circ \healthy{CSP1}(ok' \land \lnot wait' \land x' = e \land dtr' = dtr \land v' = v)$ \\[+1ex]
\textbf{Prefix (2).} In the basic model, the occurrence of an action $a$ corresponds with the predicate $do_{\mathcal{A}}(a)$. For an alphabetised event $(s,e)$ we want to record the dynamic interface $s$ as well as the event $e$. The value of $s$ may be given by the value of $mChans$ at the time of the record. The process that is ready to engage in event $a$ and then increments the DAT when $a$ has occurred, or simply records the current dynamic interface (to serve as the valid interface for the next process) is denoted by $mdo_{\Sigma}(a)$.
\begin{definition} [$mdo_{\Sigma}(a)$] 
\negbigskip%
\begin{align*}
& mdo_{\Sigma}(nil) \circdef dtr := dtr \cat \langle (mChans, nil) \rangle  \\%
& \emph{For any event } a \neq nil:  \\%
& mdo_{\Sigma}(a) \circdef m\Phi (a \notin dref' \lhd wait' \rhd dtr' = dtr \cat \langle (mChans, a) \rangle)  \\%
& \emph{where } \Phi~ \text{is defined as in def.\ref{df:phi-do} and } m\Phi \circdef \mathbf{MC123} \circ \Phi.
\end{align*}
\end{definition}
\negbigskip\negbigskip%
\subsection{Channel-passing} \negbigskip%
Moving a channel has different effects depending on whether the channel is being moved out (released) or moved in (acquired). \\[+1ex]
\textbf{Release.} Moving out/sending out a channel implies that the channel must no longer be authorised i.e.\ it must be removed from $mChans$. Clearly, any attempt of moving out a channel, say $oldch$, not already owned must fail: formally, this gives the assumption $(oldch \in mChans)_{\bot}$. Because of $\mathbf{MC2}$, the new value of $mChans$ must be recorded into the trace i.e.\ $mdo_{\Sigma}(nil)$, so that any future refusal may not contain the event that has just been removed. This further means that all of the events related with the channel must be removed from $dref$ as well, if they were already in $dref$, to avoid chaotic behaviour. The operation for releasing a channel will be called (channel) \emph{s-assignment} and denoted by $:=_{s}$.
\begin{definition} [Channel s-assignment]
Let $oldch$ be the channel to be released,then: 
\begin{align*}
  \begin{aligned}
  (\kappa ch :=_{s} oldch) \circdef 
    &\left( \begin{aligned}  
   &(oldch \in mChans)_{\bot} \comp  & \\[-1ex]%
    &\left(
      \begin{aligned}
       \kappa ch \\[-1ex] mChans \\[-1ex]  ref
      \end{aligned}  \right) 
      :=  
      \left( \begin{aligned}
          &oldch \\[-1ex] &mChans \setminus \{oldch\} \\[-1ex]%
      & \left(
        \begin{aligned}
          &ref \setminus \alpha~oldch \\[-1ex]%
          &\lhd (\alpha~oldch \in ref) \rhd ref
        \end{aligned} \right)
      \end{aligned} \right)      
      \comp \\[-1ex]%
    & mdo_{\Sigma}(nil) 
    \end{aligned} \right)
  \end{aligned}
\end{align*}
To model the situation where the sending process just sends the channel but still retains its value, normal or clone assignment ($:=$) may be used. 
\end{definition}
Any attempt of using a channel after it has been moved out by s-assignment leads to undefinedness or $CHAOS$. This further guarantees that a channel may not send itself (s-assignment removes it from the interface).
\begin{theorm} [Undefined channel]
$(\kappa chans :=_{s} ch \comp ch.e) = CHAOS$
\end{theorm}
\textbf{Acquisition.} Moving in /receiving a channel $newch$ requires that the receiving process must not own $newch$ prior to receiving it, which corresponds to the assumption $(newch \notin mChans)_{\bot}$. $newch$ must then be added into $mChans$. Due to $\mathbf{MC2}$, the value of $mChans$ must be recorded into the trace, which will notably permit to increment the value of $dref$ with the events of the acquired channel, subsequently. The operation for acquiring a new channel will be called (channel) \emph{r-assignment} and denoted by $:=_{r}$.
\begin{definition}[Channel r-assignment]
Let $newch$ be the channel to be acquired. Then: %
  \begin{align*}
  (\kappa ch :=_{r} newch) \circdef 
    &\left( \begin{aligned}  
   &(newch \notin mChans)_{\bot} \comp  & \\[-1ex]%
    &\left(
      \begin{aligned}
       &\kappa ch \\[-1ex] &mChans
      \end{aligned}  \right) 
      :=  
      \left( 
        \begin{aligned}
          &newch \\[-1ex] &mChans \cup \{newch\} \\[-1ex]  
        \end{aligned} \right)      
      \comp \\[-1ex]%
    & mdo_{\Sigma}(nil) 
    \end{aligned} \right)
  \end{align*}
To model the situation where the sending process just sends the channel but still retains its value, normal or clone assignment ($:=$) may be used.%
\end{definition}
The preceding definition actually states that the behaviour of the process in the case where it receives a channel already owned should be $CHAOS$. That is a quite strict definition but it is up to the programmer to implement that behaviour however he would like to, probably though, by throwing an exception. \\%
To further ensure that it is r-assignment $:=_{r}$ and not just assignment $:=$ that is used when receiving a channel, we define a new input prefix denoted by $ch??$, which behaves the same as $ch?$ except that normal assignment $:=$ is replaced by r-assignment $:=_{r}$.
\begin{definition}[Channel-passing input prefix] ~~~\\
\indent
$(ch1?? \kappa chans \circthen P) \circdef \Extchoice_{newch} mdo_{\Sigma}(ch1.newch) \comp \kappa chans :=_{r} newch \comp P $
\end{definition}
Channel mobility is characterised by s-assignment for the sending/source process, and by r-assignment for the receiving/target process. 
\negbigskip\negbigskip\negbigskip%
\section{ Discussion and Related Work}\label{mobC:relwork}\negbigskip\negbigskip%
\subsection{A new concept: the \emph{capability} of a process}\label{disc:capa} 
\negbigskip%
We have introduced at the beginning of section \ref{mobC:concepts} the set $Chans$ of concrete static channels and its equivalent $MCh$ of mobile channels. Although $Chan$ was quickly discarded, its relation to the static interface viz.\ $Chans = \mathcal{I}$ was inherited between $MCh$ and the dynamic interface. (Although we shall discuss the case of $Chans$ only, the following applies equally to $MCh$.) The question that may be raised is this: why introduce $Chans$ if only to discard it, by equating it to the interface? Why not use the interface straight from the beginning? \\%
This question has a great theoretical interest. The difficulty of working with interfaces is in their very meaning and use. Traditionally, an interface means \emph{authorised} channels, namely those that a process is allowed to use, or equivalently, those that a process \emph{owns}. Since a process owns them at all times, it may or not use them. In other words, it is enough to state at what time a channel may be used to obtain a somewhat dynamic effect. Yet, if one wants to say that ownership may change, the question quickly arises: whence come the new channels? If we say that they were already in the interface, it would seem that the only thing that we have achieved is sending a signal to state when to actually (in time) use what channel. Clearly, no one can perceive mobility in such a device.  \\%
On the other hand, since channels are only used for communication, they play no important role of themselves. They are only interesting insofar as they provide a nice means for constructing (communication) events. In fact, not everyone uses interfaces systematically. The alphabetised parallel operator \cite[\S 2.2]{Ros98} is a symptom of this: we only specify interfaces for the sake of synchronisation. Hence in practice, it is more current to derive (or compute) the interface from the set of events rather than the contrary. \\%
In the case of mobility, since channels are manipulated explicitly, they must in consequence be treated like concrete entities (see e.g.\ \cite{GroSto96b}, \cite{BialPes09b}). Fortunately, since the effect of channel mobility may be readily discussed in terms of its effect on the interface, the notion of concrete channels is quickly dissolved into the interface as well. \emph{This is the first work where it is clearly shown how the two notions coincide.} \\%
Indeed, the equality $Chans = \mathcal{I}$ is an equality on names, and not that of meaning. For a clearer distinction, we call both $Chans$ and $MCh$ the \emph{capability} of a process: it represents the capability, or the capacity for a process to acquire a new channel. This does not imply the ownership of said channels, properly defined by the process's \emph{interface}. Unlike the interface that may change, the capability may not; it is its own maximal value. \\%
An analogy may be drawn in the relation between the capability of a child to grow in knowledge and the actual knowledge of the child. The capability is in the child, yet not the knowledge; as the child grows in knowledge, he realises his capability of growing in knowledge; hence he knows of his capability of growing in knowledge. He may lose the knowledge of his capability (seeing how he never knew it initially), yet he may never lose the capability itself. He cannot lose the latter since by losing it, he would either realise his loss (thus he has gained more knowledge, a contradiction), or he would not (which is impossible since he has the capability of knowing before the loss). Conversely, since by his capability of knowing he comes to know of his capability, it comes that by the capability of knowing, he also has the capability of knowing his capability. Thus there can be no increase of his capability of knowledge. \\%
Another analogy may be drawn with the mathematical concept of the proof of the existence of limits. That one has proved that the limit of a function exists says nothing about the actual value of the limit. A difference though, is that here, the existence of the capability is defined axiomatically. The knowledge (ownership) of a channel is determined by evaluation of the interface. 
\negbigskip\negbigskip%
\subsection{Related Work} \negbigskip%
%
In \cite{Broy95}, .., \cite{Berg99}, Broy et al.\ consider an extension of FOCUS semantics framework \cite{Broy93} with mobility. In FOCUS, programs are formally represented as (sets of) functions from input histories to output histories. 
Grosu and Stolen \cite{GroSto96a, GroSto96b, GSB97, GroSto99}, based on a timed nondeterministic model, characterise channel mobility by allowing functions to change of history (input and output), in time. %
They define a \emph{privacy preservation} law that plays the same role as our healthiness condition $\healthy{MC1}$. Since they consider that such their privacy law may not be stated in an untimed model, our results seems to contradict with theirs. In fact, a careful analysis of their results shows that what is actually specified in time are the valid interfaces, as can be seen from \cite[Section 4]{GroSto96b}. %
Hence, if we replace clock ticks in histories by the actual interface at corresponding times, we would obtain an untimed model for channel mobility in FOCUS. \\[+1ex]%
In \cite{WelBa08} Welch and Barnes give a CSP denotational semantics to the channel mobility mechanism of occam-pi programming language \cite{WelBa04}. The model contains too many implementation details, however. We have found that said model may actually be simplified without much difficulty. Seeing how mobile channels are modelled as (indexed) processes, and yet the processes themselves do not move, it is enough to keep the channels (with their corresponding index), and discard the processes. Hence, we obtain a simpler model by replacing indexed processes with sets of indexed channels viz.\ we lift the indexing procedure into a channel-naming procedure. Then, by modelling indexed channels as mobile channels (since indexed channels are but channels) using our model, we obtain a (abstract) model for occam-pi' channel mobility mechanism. \\[+1ex] 
In \cite{SchTre07, Vaj09}, an extension of CSP$||$B with channel mobility is proposed. The model is quite restrictive, since only the channels that link CSP controllers with B machines may be moved, and not the links between CSP processes themselves. Only a trace model is proposed, and the language studied does not have the hiding operator. \\[+1ex]%
In \cite{BialPes09b} Bialkiewicz and Peschanski define a trace model for pi-calculus processes using the notion of localised traces, with the aim of achieving a CSP-like trace model. Much effort is dedicated to achieving name freshness similar to the pi-calculus. The resulting model demands a complex manipulation of locations and is in general less expressive than CSP. The use of locations further limits the possibility of extending  the model with failures and divergences. \\[+1ex]%
%
In \cite{Ros10b} Roscoe shows that pi-calculus operators are CSP-like, though this does not equate CSP with a mobility language. He further introduces possible models for the pi-calculus in CSP. 
In (\cite{Ros10}, \S 20.3), Roscoe discusses a way of adding mobility into CSP directly, without referring to the pi-calculus. The semantics are built for a restricted type of mobility where the set of channels to be moved is known in advance: they are also called \emph{closed-world} semantics. In particular, both events and channels may be moved. We strongly believe that event mobility may not be used for describing channel mobility, for the very difference between events and channels. Also, we may argue that the difference between open and closed world lies in the eye of the observer. Since a process always knows what channels it owns, it needs only to care about them. Hiding is the job of another process, which needs to care only about hiding. Hence, as in our semantics, the process resolves name clashes. However, this does not mean that information may leak. Indeed, everyone knows that computers have internal buses, but no one except the technician may access those. Hence, to even send a channel name to a process requires having the right sending channel to begin with: security is thus ensured by attributing channels accordingly. \\[+1ex]%
In \cite{HoaOh08} Hoare and O'Hearn have defined semantics for channel mobility combining ideas from both Separation Logic and CSP. Their model is similar to ours, despite some differences. In general though, our work may be considered to extend theirs with failures and divergences. Hoare and O'Hearn have also defined allocation and deallocation operations which may change (resp.\ increase and decrease) the interface of a process, without the participation of the environment. %
This permits drawing a subtle difference in our characterisation of alphabets, with that of Hoare and O'Hearn. %
Hoare and O'Hearn, not having the notion of a (process) \emph{capability} (cf.\ \S \ref{disc:capa}), do not tell where allocated channels come from. If they were already in the interface, then clearly, no mobility has occurred at all; and if they were not then they seem to have been created ex nihilo. Our notion of capability clearly discards creation ex nihilo, which would lead to chaotic behaviour: if a process can decide what a channel is and what is not, it may as well decide contrary to the expectations of the programmer. In our model, allocation would be modelled differently: by means of a communication with the environment, followed by hiding of the channel used for the acquisition. %
\paragraph{Acknowledgements.} The author is grateful to Pr.\ Jim Woodcock for his advices and numerous comments and suggestions about the structure and contents of this paper. Without his help this paper would not have been. Thanks also to the anonymous reviewers for their useful comments.
\negbigskip\negbigskip\negbigskip%
\section{Conclusion}\negbigskip
In this paper we have presented preliminary results on work aimed at defining the semantics for channel mobility in UTP-CSP. The model obtained retains the simplicity of the original CSP model and allows for their comparison. To date, we have defined the semantics of many operators, especially parallel composition and hiding, though that work has not been published yet. We have also defined links between static and mobile UTP-CSP and the result is promising. We hope to complete the remaining work soon. %
The question of the link between CSP and the pi-calculus is also an interesting area of future research. 
\negbigskip\negbigskip\negbigskip%


\begin{thebibliography}{9} \negbigskip%
\bibitem{Mil99} R. Milner, \emph{Communicating and Mobile Systems: the pi-calculus}, Cambridge University Press, 1999.

\negbigskip%
\bibitem{HoaHe98} T. Hoare, Jifeng He, \emph{Unifying Theories of Programming}, Prentice-Hall, 1998.

\negbigskip%
\bibitem{CavWoo06} Ana Cavalcanti, J. Woodcock, \emph{A Tutorial Introduction to CSP in Unifying Theories of Programming}, Refinement Techn. in Softw. Eng., pp. 220-268, Springer, 2006. \doi{10.1007/11889229\_6}

\negbigskip%
\bibitem{Hoa85} T. Hoare, \emph{Communicating Sequential Processes}, Prentice-Hall, 1985.

\negbigskip%
\bibitem{Ros98} A.W. Roscoe, \emph{The Theory and Practice of Concurrency}, Prentice-Hall, 1998.

\negbigskip%
\bibitem{Ros10} A.W. Roscoe, \emph{Understanding Concurrent Systems}, Prentice-Hall, 2010. \doi{10.1007/978-1-84882-258-0}

\negbigskip%
\bibitem{Ros10a} A.W. Roscoe, \emph{On the expressiveness of CSP}, 2011 draft, Available at \href{http://www.cs.ox.ac.uk/files/1383/expressive.pdf}{cs.ox.ac.uk/ros11.pdf}

\negbigskip%
\bibitem{Ros10b} A.W. Roscoe, \emph{CSP is expressive enough for pi}, In `Reflections on the Work of C.A.R. Hoare', History of Computing 2010, pp. 371-404, 2010. \doi{10.1007/978-1-84882-912-1\_16}

\negbigskip%
\bibitem{WelBa04} P.H. Welch, Frederick R.M. Barnes, \emph{Communicating Mobile Processes - Introducing occam-pi}, Lecture Notes In Comp. Sci.(LNCS), vol.3525, pp. 175-210, Springer, 2004. \doi{10.1007/11423348\_10}

\negbigskip%
\bibitem{WelBa08} P.H. Welch and F.R.M. Barnes, \emph{A CSP Model for Mobile Channels}, Communicating Process Architectures(CPA), vol.66, pp.17-33, IOS Press, 2008. \doi{10.3233/978-1-58603-907-3-17}

\negbigskip%
\bibitem{Pes04} F. Peschanski, \emph{On Linear Time and Congruence in Channel-Passing Calculi}, CPA, pp. 39-54, IOS Press, 2004.

\negbigskip%
\bibitem{BialPes09a} J-A. Bialkiewicz, F. Peschanski, \emph{Logic for Mobility: A Denotational Approach}, Logic, Agents and Mobility(LAM), pp. 44-59, 2009. Available at \href{http://www.wotug.org/papers/CPA-2009/BialkiewiczPeschanski09/BialkiewiczPeschanski09.pdf}{wotug/CPA09/BP09.pdf}

\negbigskip%
\bibitem{BialPes09b} J-A. Bialkiewicz, F. Peschanski, \emph{A Denotational Study of Mobility}, CPA, pp. 239-261, 2009. \doi{10.3233/978-1-60750-065-0-239}

\negbigskip%
\bibitem{Sta96} I. Stark, \emph{A fully-abstract domain model for the pi-calculus}, Logic in Comp. Sci.(LICS), pp. 36-42, 1996. \doi{10.1109/LICS.1996.561301}

\negbigskip%
\bibitem{Fio96} M. Fiore, E. Moggi, and D. Sangiorgi, \emph{A fully-abstract model for the pi-calculus}, LICS, pp. 43-54, IEEE, 1996. \doi{10.1109/LICS.1996.561302}

\negbigskip%
\bibitem{Hen02} M. Hennessy, \emph{A fully-abstract denotational semantics for the pi-calculus}, Theoretical Computer Science (TCS), vol.278, pp. 53-89, Elsevier, 2002. \doi{10.1016/S0304-3975(00)00331-5}

\negbigskip%
\bibitem{HoaOh08} T. Hoare, P. O’Hearn, \emph{Separation Logic Semantics for Communicating Processes}, LNCS, pp. 3-25, Elsevier, 2008. \doi{10.1016/j.entcs.2008.04.050}

\negbigskip%
\bibitem{Broy93} M. Broy, F. Dederichs, C. Dendorfer, M. Fuchs, T. F. Gritzner, and R. Weber, \emph{The design of distributed systems -- an introduction to FOCUS}, Tech. Rep., Uni. of Munich, January 1992. 

\negbigskip%
\bibitem{Broy95} M. Broy, \emph{Equations for Describing Dynamic Nets of Communicating Systems}, Recent Trends in Data Type Specification,  pp. 170-187, Springer, 1995. \doi{10.1007/BFb0014427} 

\negbigskip%
\bibitem{Broy14}  M. Broy, \emph{A Model of Dynamic Systems}, LNCS, vol.8415, pp.39-53, Springer, 2014. \doi{10.1007/978-3-642-54848-2\_3}

\negbigskip%
\bibitem{GroSto96a} R. Grosu, K. Stolen, \emph{A Denotational Model for Mobile P2P DFNs without Channel Sharing}, Tech. Rep., Uni. of Munich, Sep.1996. 

\negbigskip%
\bibitem{GroSto96b} R. Grosu, K. Stolen, \emph{Specification of Dynamic Networks}, Tech. Rep., Uni. of Munich, Dec.1996.

\negbigskip%
\bibitem{GSB97} R. Grosu, K. Stolen, M. Broy, \emph{A Denotational Model for Mobile P2P Data Flow Nets with Channel Sharing}, Tech. Rep., Uni. of Munich, May.1997.

\negbigskip%
\bibitem{GroSto99} R. Grosu, K. Stolen, \emph{Stream based specification of Mobile Systems}, Formal Aspects of Computing, vol. 13, pp. 1-31, Springer, 2001. \doi{10.1007/PL00003937}

\negbigskip%
\bibitem{Sto99}  K. Stolen, \emph{Specification of Dynamic Reconfiguration in the Context of Input/Output Relations}, In Formal Methods for Open Object-Based Distributed Systems, pp. 259-272, Springer, 1999. \doi{10.1007/978-0-387-35562-7\_20}

\negbigskip%
\bibitem{Berg99}  K. Bergner, R. Grosu, A. Rausch, A.Schmidt, P. Scholz, M. Broy, \emph{Focusing on Mobility}, Proc. of the 32nd Hawaii Internat. Conf. on Sys. Sci., IEEE, 1999. \doi{10.1109/HICSS.1999.773061}

\negbigskip%
\bibitem{SchTre07} S. Schneider, H. Treharne, and B. Vajar, \emph{Introducing mobility into CSP$||$B}, In Automated Verification of Critical Systems (AVoCS), 2007.

\negbigskip%
\bibitem{Vaj09} Beeta Vajar, \emph{Mobile CSP$||$B}, AVoCS, 2009.
%
\end{thebibliography}
\end{document}